\documentclass[twocolumn]{svjour3}
\usepackage{graphics}
\usepackage{amsmath}

\begin{document}

\title{A Nonlocal Wave-Particle Duality}
\author{Mohd Asad Siddiqui \and Tabish Qureshi}
\institute{Centre for Theoretical Physics, Jamia Millia Islamia, New Delhi,
India. \email{tabish@ctp-jamia.res.in}}

\maketitle


\begin{abstract}
We propose and analyse a modified ghost-interference experiment, and show that
revealing the particle-nature of a particle passing through a double-slit
hides the wave-nature of a spatially separated particle which it is
entangled with. We derive a {\em nonlocal duality relation},
${\mathcal D}_1^2 + {\mathcal V}_2^2 \le 1$, which connects
the path distinguishability of one particle to the interference visibility
of the other. It extends Bohr's principle of complementarity to a nonlocal 
scenario.  We also propose a 
{\em ghost quantum eraser} in which, erasing the which-path information of
one particle brings back the interference fringes of the other.
\end{abstract}


\section{Introduction}

The two-slit interference experiment has become a cornerstone of
the issue of wave-particle duality and Bohr's complementarity principle.
So beautifully and simply does it
capture the dual nature of particles and light and the superposition
principle that it has become symbolic of the mysterious nature of quantum mechanics.
The fact that the wave and particle nature cannot be observed at the same time,
appears to be so fundamental that Bohr elevated it to the level of a new
principle, the principle of complementarity\cite{bohr}. Bohr asserted that
if an experiment clearly revealed the particle nature, it would completely
hide the wave nature, and vice-versa. This principle has stood its ground
in face of several attacks over the years.

This principle has now been made quantitatively precise by a
bound on the extent to which the two natures could be simultaneously observed
\cite{greenberger,englert}. The extent to which one 
can distinguish which of the two slits a particle passes through, is given by
a quantity ${\mathcal D}$, called the path-distinguishability, quantifying
the particle nature. The wave-nature is quantified by the visibility of the
interference pattern, given by
${\mathcal V}$. The quantities ${\mathcal D}$ and ${\mathcal V}$ are 
so defined that they can take values only between 0 and 1. The relation
putting a bound on the two is given by the so-called
duality relation\cite{englert}
\begin{equation}
{\mathcal V}^2 + {\mathcal D}^2 \le 1.
\label{egy}
\end{equation}
The above relation implies that a full which-path 
information (${\mathcal D}=1$) would definitely wash out the interference
pattern completely (${\mathcal V}=0$).

It is quite obvious that when
we talk of path distinguishability, we talk of the which-path 
knowledge of the same particle which eventually contributes to the interference
pattern. In this sense the duality relation (\ref{egy}) is {\em local}. In the following
we propose and theoretically analyse an experiment involving pairs
of entangled particles in which we relate the which-path information
of one particle to the fringe visibility of the other.

\section{Ghost interference}

The starting point of our analysis is the well known ghost-interference
experiment carried out by Strekalov et al.\cite{ghostexpt}. In this
experiment, pairs of entangled photons are generated from a
spontaneous parametric down conversion (SPDC) source. In the path of
photon 1 is kept a double-slit and further down in the path is a {\em fixed}
detector D1. Photon 2 travels undisturbed and is ultimately detected by 
the movable detector D2. The detectors D1 and D2 are connected to a 
coincidence counter. In coincident counts, a two-slit interference pattern
is seen by detector D2 for photon 2. Note that photon 2 does not pass through
any double-slit. This interference was appropriately called ghost
interference, and has been understood to be a consequence of entanglement.
This experiment generated lot of research attention in subsequent years
\cite{ghostimaging,rubin,zhai,jie,zeil2,pravatq,twocolor,sheebatq,ghostunder}.
The photon pairs emerging from an SPDC source are believed to capture the
essence of the EPR state \cite{epr} introduced by Einstein, Podolsky and
Rosen \cite{klyshko}. The spatial correlations shown by photons emerging
from parametric down-conversion are well understood now \cite{walborn}.
\begin{figure}
\centerline{\resizebox{8.5cm}{!}{\includegraphics{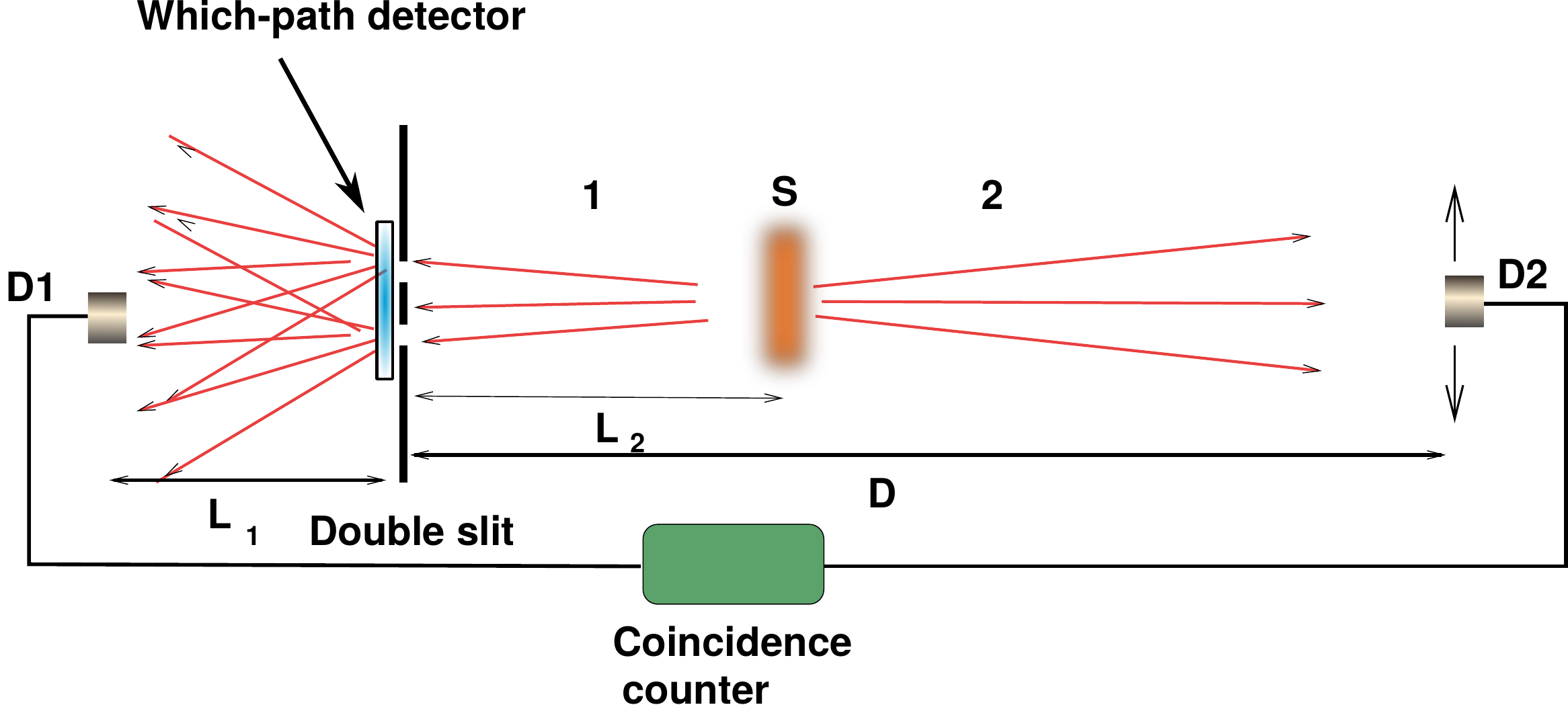}}}
\caption{A modified ghost-interference experiment with entangled pairs of
particles. A which-path detector can detect which of the two slits
particle 1 went through. Detector D1 is fixed at $z=0$, but D2 is free to
move along the z-axis.}
\label{ghostmod}
\end{figure}

Our proposed experiment is shown in Fig.\ref{ghostmod}. Entangled particle
pairs emerge from a source S. For the sake of generality, let us assume
massive particles, although the ghost interference experiment is done with
photons. Particle 1 passes through a double-slit and also interacts with
a which-path detector. We do not specify any form of the which-path detector,
but just assume that it is a quantum system, initially in a state $|d_0\rangle$.
If the particle passes through slit 1, the path detector ends up in the 
state $|d_1\rangle$, and if it passes through slit 2, the path detector
ends up in the state  $|d_2\rangle$. In general, when the particle has passed
through the double-slit, the two path detector states, $|d_1\rangle$ and
$|d_2\rangle$, get entangled with the two paths of particle 1.
This entanglement is a must in order that the which-path detector acquires
the relevant information about the particle. Particle 1 then travels and
reaches a {\em fixed} detector D1. Particle 2 travels unhindered to the 
detector D2. As the two particles have to be counted in coincidence,
the paths travelled by both the particles, before reaching their respective
detectors, are equal. Without the which-path detector, this experiment is
just the original ghost interference experiment where particle 2 displays
an interference pattern\cite{ghostexpt}.

Essentially we are now looking at the issue of wave-particle duality in
a system of entangled particles, a subject not particularly well studied.
Wave-particle duality for entangled systems has been studied in a rather
generalized formalism by Vaccaro \cite{vaccaro}.

\section{Which-path information}

We assumes $|d_1\rangle,|d_2\rangle$ to be normalized, but not necessarily
orthogonal. The ultimate limit to the knowledge we can acquire as to
which slit particle 1 went through, is set by how distinct the states
$|d_1\rangle,|d_2\rangle$  are. If $|d_1\rangle,|d_2\rangle$  are orthogonal,
we can {\em in principle} know with hundred percent accuracy which slit the
particle went through. With this thinking we define which-path
distinguishability {\em for particle 1} as
\begin{equation}
{\mathcal D}_1 = \sqrt{ 1 - |\langle d_1|d_2\rangle|^2}.
\label{duality}
\end{equation}

In order to quantify the effect of the which-path detector on the ghost
interference shown by particle 2, we carry out a quantum mechanical
analysis of the dynamics of the entangled particles. We assume that the
particles travel in opposite directions along the x-axis. The entanglement
is in the z-direction. The best state to describe momentum-entangled
particles is the {\em generalized EPR state}\cite{tqajp}
\begin{equation}
\Psi(z_1,z_2) = C\!\int_{-\infty}^\infty dp
e^{-p^2/4\hbar^2\sigma^2}e^{-ipz_2/\hbar} e^{i pz_1/\hbar}
e^{-{(z_1+z_2)^2\over 4\Omega^2}}, \label{state}
\end{equation}
where $C$ is a normalization constant, and $\sigma,\Omega$ are certain
parameters. In the limit $\sigma,\Omega\to\infty$ the state (\ref{state})
reduces to the EPR state introduced by Einstein, Podolsky and Rosen\cite{epr}.

After performing the integration over $p$, (\ref{state}) reduces to
\begin{equation}
\Psi(z_1,z_2) = \sqrt{ {\sigma\over \pi\Omega}}
 e^{-(z_1-z_2)^2\sigma^2} e^{-(z_1+z_2)^2/4\Omega^2} .
\label{psi0}
\end{equation}
It is straightforward to show that $\Omega$ and $\sigma$ quantify the position
and momentum spread of the particles in the z-direction.

We assume that after travelling for a time $t_0$, particle 1 reaches the 
double slit ($vt_0 = L_2$), and particle 2 travels a distance $L_2$ towards
detector D2.
Using the strategy outlined in the preceding discussion, we can write
the state of the entangled photons after a time $t_0$.

The state of the entangled system, after this time evolution, can be
calculated using the Hamiltonian governing the time evolution, given by
$\hat{H} = {p_1^2\over 2m} +{p_2^2\over 2m}$.
After a time $t_0$, (\ref{psi0}) assumes the form
\begin{eqnarray}
\Psi(z_1,z_2,t_0) = C_{t_0}
\exp\left[{-(z_1-z_2)^2\over {1\over\sigma^2} + {4i\hbar t_0\over m}}
-
{(z_1+z_2)^2\over 4\Omega^2  +
{i\hbar t_0\over m}} \right],\nonumber\\
\label{Psit}
\end{eqnarray}
where $C_{t_0}=\left({\pi}(\Omega+{i\hbar t_0\over 4m\Omega})
(1/\sigma + {4i\hbar t_0\over m/\sigma})\right)^{-1/2}$.

We take into account the effect of the double-slit on the 
entangled state as follows. We assume that the double-slit allows
the portions of the wave-function in front of the slit to pass through,
and blocks the other portions. We assume that what
emerge from the double-slit are localized Gaussian wavepackets, whose
width is the width of the slit. The two slits being A and B, the wavepackets
which pass through, are denoted by $|\phi_A(z_1)\rangle$ and
$|\phi_B(z_1)\rangle$,
respectively. The portion of particle 1 which gets blocked is, say,
$|\chi(z_1)\rangle$. These three states are obviously orthogonal, and the
entangled two-particle state can be expanded in terms of these.
We can thus write:
\begin{equation}
|\Psi(z_1,z_2,t_0)\rangle = |\phi_A\rangle|\psi_A\rangle
+ |\phi_B\rangle|\psi_B\rangle +
|\chi\rangle|\psi_{\chi}\rangle , \label{slit}
\end{equation}
where the corresponding states of particle 2 are given by
\begin{eqnarray}
|\psi_A(z_2)\rangle &=& \langle\phi_A(z_1)|\Psi(z_1,z_2,t_0)\rangle \nonumber\\
|\psi_B(z_2)\rangle &=& \langle\phi_B(z_1)|\Psi(z_1,z_2,t_0)\rangle \nonumber\\
|\psi_\chi(z_2)\rangle &=& \langle\chi(z_1)|\Psi(z_1,z_2,t_0)\rangle \label{psi}.
\end{eqnarray}
In addition, the wavepackets of particle 1 get entangled with the two
states of the which-path detector $|d_1\rangle,|d_2\rangle$.
So, the state we get after particle 1 crosses the double-slit is:
\begin{equation}
|\Psi(z_1,z_2)\rangle = |d_1\rangle|\phi_A\rangle|\psi_A\rangle
+ |d_2\rangle|\phi_B\rangle|\psi_B\rangle +
|d_0\rangle|\chi\rangle|\Psi_\chi\rangle .
\end{equation}
The first two terms represent the amplitudes of particle 1 passing through
the double-slit, and the last term represents the amplitude of it getting
blocked. If the particle 1 gets blocked, the state of the path detector
remains what it was initially, i.e., $|d_0\rangle$. Unitarity of the
dynamics makes sure that these three parts of
the wave-function will evolve independently, without affecting each
other. Since we are only interested in situations where particle 1 passes
through the slit, we will throw away the term which represents
particle 1 not passing through the slits. To do that, we just have to
renormalise the remaining part of the wave-function, which looks like
\begin{equation}
|\Psi(z_1,z_2)\rangle = {1\over A} (|d_1\rangle|\phi_A\rangle|\psi_A\rangle
+ |d_2\rangle|\phi_B\rangle|\psi_B\rangle),
\end{equation}
where $A = \sqrt{\langle\psi_A|\psi_A\rangle + \langle\psi_B|\psi_B\rangle}$.

In the following, we assume that $|\phi_A\rangle$, $|\phi_B\rangle$, are
Gaussian wave-packets:
\begin{eqnarray}
\phi_A(z_1) &=& {1\over(\pi/2)^{1/4}\sqrt{\epsilon}} e^{-(z_1-z_0)^2/\epsilon^2}
,\nonumber\\
\phi_B(z_1) &=& {1\over(\pi/2)^{1/4}\sqrt{\epsilon}} e^{-(z_1+z_0)^2/\epsilon^2}
,
\label{gaussians}
\end{eqnarray}
where $\pm z_0$ is the z-position of slit A and B, respectively, and $\epsilon$
their widths. Thus, the distance between the two slits is $2 z_0 \equiv d$.

Using (\ref{Psit}), (\ref{psi}) and (\ref{gaussians}), wave-functions $|\psi_A\rangle$,
$|\psi_B\rangle$ can be calculated, which, after normalization, have the form
\begin{equation}
\psi_A(z_2) = C_2 e^{-{(z_2 - z_0')^2 \over \Gamma}},~~~
\psi_B(z_2) = C_2 e^{-{(z_2 + z_0')^2 \over \Gamma}} ,
\end{equation}
where $C_2 = (2/\pi)^{1/4}(\sqrt{\Gamma_R} + {i\Gamma_I\over\sqrt{\Gamma_R}})^{-1/2}$,
\begin{equation}
z_0' = {z_0 \over {4\Omega^2\sigma^2+1\over
4\Omega^2\sigma^2-1} + {4\epsilon^2 \over 4\Omega^2-1/\sigma^2}},
\end{equation}
and
\begin{equation}
\Gamma = \frac{\epsilon^2+ {1\over\sigma^2}+{\epsilon^2\over 4\Omega^2\sigma^2}
 + {2i\hbar t_0\over m} 
}{1 + {\epsilon^2\over\Omega^2}+{i2\hbar t_0\over 4\Omega^2m} +
{1\over 4\Omega^2\sigma^2}} + {2i\hbar t_0\over m}.
\end{equation}
Thus, the state which emerges from the double slit, has the following form
\begin{eqnarray}
\Psi(z_1,z_2) &=& c|d_1\rangle e^{{-(z_1-z_0)^2\over\epsilon^2}}
e^{{-(z_2 - z_0')^2 \over \Gamma}} \nonumber\\
&&+ c|d_2\rangle e^{{-(z_1+z_0)^2\over\epsilon^2}}
e^{{-(z_2 + z_0')^2 \over \Gamma}} \label{virtual},
\end{eqnarray}
where $c = (1/\sqrt{\pi\epsilon})(\sqrt{\Gamma_R} +
{i\Gamma_I\over\sqrt{\Gamma_R}})^{-1/2}$. 
Particles travel for another time $t$ before reaching their respective 
detectors. We assume that the wave-packets travel in the x-direction with
a velocity $v$ such that $\lambda=h/mv$ is the de Broglie wavelength.
Using this strategy, we can
write $\hbar (t+2t_0)/m = \lambda D/2\pi$, $\hbar t_0/m = \lambda L_2/2\pi$.
The expression $\lambda D/2\pi$ will also hold for a
photon provided, one uses the wavelength of the photon for
$\lambda$\cite{ghostunder}. 
The state acquires the form
\begin{eqnarray}
\Psi(z_1,z_2,t)=C_t
|d_1\rangle\exp\left[{{-(z_1-z_0)^2\over\epsilon^2+{iL_1\lambda\over\pi}}}\right]
 \exp\left[{{-(z_2 - z_0')^2 \over \Gamma+{iL_1\lambda\over\pi}}}\right]
\nonumber\\
+ C_t |d_2\rangle\exp\left[{{-(z_1+z_0)^2\over\epsilon^2+{iL_1\lambda\over\pi}}}\right]
 \exp\left[{{-(z_2 + z_0')^2 \over \Gamma+{iL_1\lambda\over\pi}}}\right],\nonumber\\
\label{psifinal}
\end{eqnarray}
where 
\begin{equation}
C_t = {1\over \sqrt{\pi}\sqrt{\epsilon+iL_1\lambda/\epsilon\pi}
\sqrt{\sqrt{\Gamma_r}+(\Gamma_i+iL_1\lambda/\pi)/\sqrt{\Gamma_r}}}.
\end{equation}

In order to get simplified results, we 
consider the limit $\Omega \gg \epsilon$ and
$\Omega \gg 1/\sigma$. In this limit
\begin{equation}
\Gamma^2 \approx \gamma^2 + 4i\hbar t_0/m ,
\end{equation}
where $\gamma^2 = \epsilon^2 + 1/\sigma^2$ and $z_0' \approx z_0$.

\section{Nonlocal wave-particle duality}

We are now in a position to calculate the probability of D2 detecting 
particle 2 at a position $z_2$, provided that D1, which is fixed at
$z_1=0$, detects particle 1.  This probability density is given by
$|\Psi(0,z_2,t)|^2$, which has the following form
\begin{eqnarray}
|\Psi(0,z_2,t)|^2 &=& \alpha
e^{{-2(z_2^2+z_0^2)\over\gamma^2+[{\lambda D\over\pi\gamma}]^2}}
\cosh\left[{4z_2z_0\over\gamma^2+[{\lambda D\over\pi\gamma}]^2}\right]\nonumber\\
&&\times\left\{ 1 + |\langle d_1|d_2\rangle|
{\cos\left[{4z_2z_0\lambda D\pi\over\gamma^4\pi^2+\lambda^2 D^2}\right]
\over \cosh\left[{4z_2z_0\over\gamma^2+[{\lambda D\over\pi\gamma}]^2}\right]}
\right\},
\label{pattern}
\end{eqnarray}
where $\alpha = |C_t|^2 e^{{-2z_0^2\over\epsilon^2+[{\lambda L\over\pi\epsilon}]^2}}$.
Eqn. (\ref{pattern}) represents a ghost interference pattern for particle 2,
even though it has not passed through any double-slit.

We can calculate the fringe visibility of the interference formed by 
particle 2. Fringe visibility is defined as 
${\mathcal V} = {I_{max}-I_{min}\over I_{max}+I_{min}}$, where
$I_{max},I_{min}$ is the maximum and minimum intensity in neighboring region
of the interference pattern\cite{born}.
The fringe visibility for particle 2, from (\ref{pattern}), is given by
\begin{equation}
{\mathcal V_2} = {|\langle d_1|d_2\rangle|\over
\cosh({4z_2z_0\over\gamma^2+[{\lambda D\over\pi\gamma}]^2})} .
\end{equation}
The visibility of ghost interference has also been derived earlier by
Barbosa \cite{barbosa}, calculating a fourth order correlation function
in the theory of Mandel and coworkers \cite{mandel}. However, a connection
of the visibility of ghost interference to the which-path information for
photon 1 has not been studied before.

As $\cosh(\theta) \ge 1$, we can write the inequality
\begin{equation}
{\mathcal V_2} \le |\langle d_1|d_2\rangle|
\end{equation}
Using (\ref{duality}), the above inequality yields
\begin{equation}
{\mathcal D}_1^2 + {\mathcal V}_2^2 \le 1
\label{gduality}
\end{equation}

The inequality (\ref{gduality}) is a very interesting one. It puts a bound on
how much which-path information for particle 1 and visibility of interference
fringes for particle 2 we can get at the same time. Clearly, full which-path
information for particle 1 implies that the interference pattern of particle 2
will be completely washed out.

Bohr's complementarity principle is made quantitatively precise by the 
inequality (\ref{egy}). However, here we have a curious scenario where
complementarity is governing two separated particles which are not even
interacting with each other. By virtue of entanglement, their natures are
also entwined with each other. Revealing the particle nature of one,
hides the wave nature of the other! It appears that in this kind of entangled
state, the two particle can either reveal their wave-nature together, or
particle nature together.

\section{Ghost quantum eraser}

In the preceding section we saw that extracting which-path information in
particle 1, leads to disappearance of interference in particle 2.
For a conventional two-slit experiment it is well known that
if we devise a way to erase the which-path information,
it is possible to recover the lost interference. This phenomenon goes
by the name of {\em quantum eraser}\cite{jaynes,scully}.

\begin{figure}
\centerline{\resizebox{8.5cm}{!}{\includegraphics{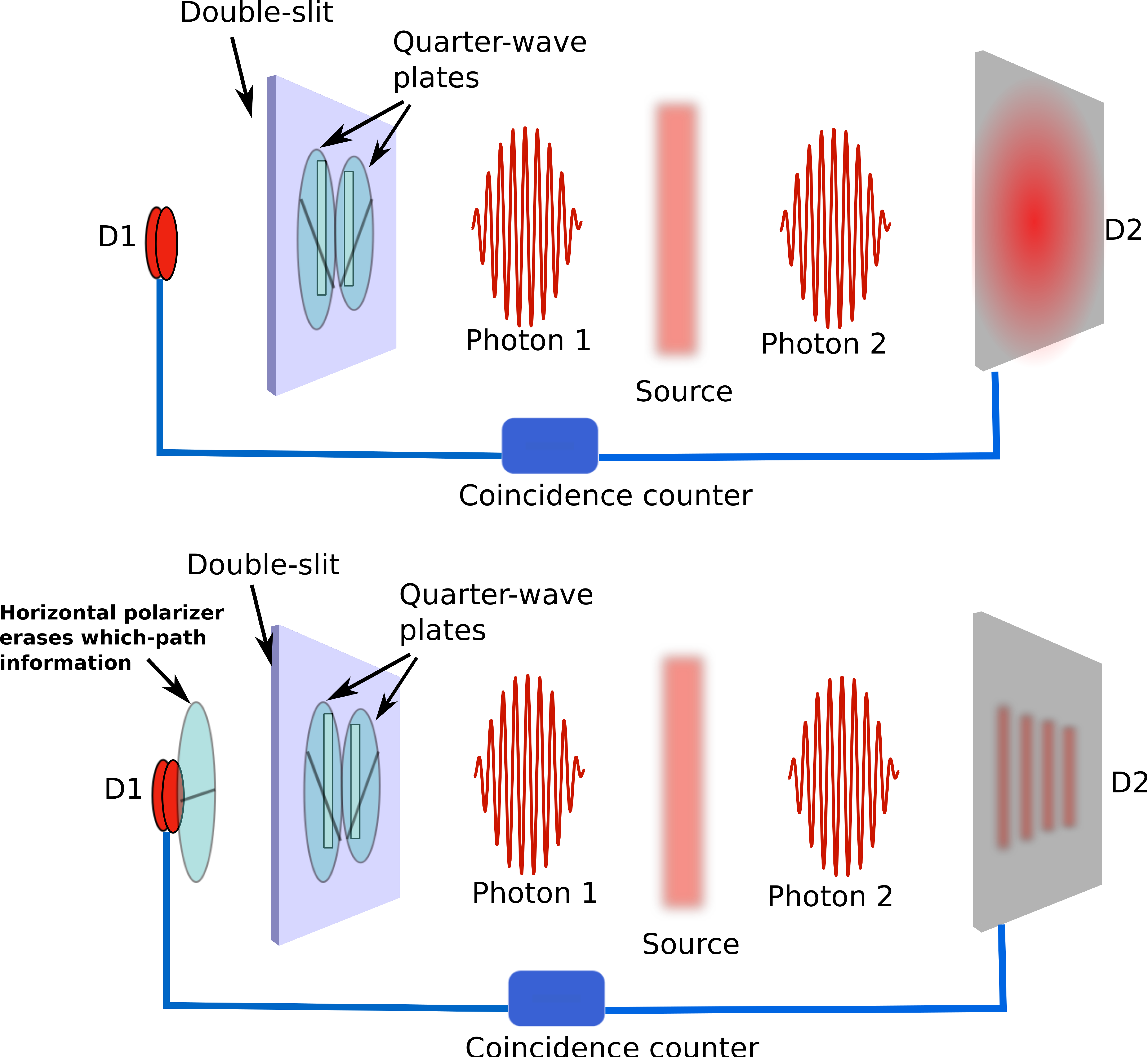}}}
\caption{Schematic diagram for the proposed {\em ghost quantum eraser.}
Quarter-wave plates in front of the slits encode which-path information
into polarization. Horizontal polarizer in front of D1 erases the which-path
information of photon 1, and the interference in photon 2 comes back.
}
\label{geraser}
\end{figure}

In the following we propose a quantum eraser experiment which
can be performed with entangled photons. The setup is shown in
Fig.(\ref{geraser}), and is motivated by a two-slit
quantum eraser demonstrated by Walborn et al.\cite{walborn1,walborn2}.
Properties of entangled photons have been used to construct 
quantum eraser earlier \cite{kim,scarcelli}. Recently quantum eraser has been
demonstrated even with thermal light \cite{peng}. Our proposal is radically different
from all these in that the which-path information and interference is 
probed, not in the same photon, but in two different photons.
The setup consists of a type I SPDC source generating pairs of
photons which we call 1 and 2. Type I SPDC source produces pairs of photons
which have the same polarization. The source is assumed to include a polarizer
which produces photon pairs which are linearly polarized.

The experiment may also be carried out using a type II SPDC source which
produces linearly polarized photons with orthogonal polarizations.
The requirement is that the two photons should not be polarization-entangled.
There is a double-slit in the path
of photon 1, followed by a {\em fixed} detector D1. Photon 2 travels
undisturbed to detector D2 which scans various positions, and acts like
a screen. The two detectors are connected to a coincidence counter.
Behind the double-slit are kept two quarter-wave plates which convert
the passing linearly polarized photons to left-circular and right-circular polarizations
respectively. This makes the which-path information about photon 1 available
to an experimenter. By probing the polarization of the photon detected at D1,
the experimenter can know which slit the photon came from.

These orthogonal polarization states of photon 1 play the role of $|d_1\rangle$
and $|d_2\rangle$, as described in the preceding analysis. However,
$|d_1\rangle$ and $|d_2\rangle$ are orthogonal in this case.
The state of the two photons, when they reach their respective detectors,
is given by (\ref{psifinal}). The coincident probability of detecting photon
2 is given by (\ref{pattern}), with $\langle d_1|d_2\rangle=0$.  There
will be no interference seen in this situation, which is depicted in the
upper diagram of Fig.\ref{geraser}.

Next we put a horizontal polarizer in front of D1. This will convert
both left- and right-circularly poralized photons to horizontally
polarized ones. All the photons reaching D1 now, are horizontally polarized,
whether they came from one slit or the other. There is no way now that
an experimenter can know which slit photon 1 came from. We have to see
what will be the behavior of photon 2 in this situation, by a careful
analysis. If the left- and right-circular polarizations are represented by
$|d_1\rangle$ and $|d_2\rangle$ respectively, adding the two would give us
linear polarization. We introduce a state
$|q_1\rangle=(|d_1\rangle+|d_2\rangle)/\sqrt{2}$ which represents horizontal
polarization. The state of the photons which reach D1 (after passing through
the horizontal plate) and D2 now, is given by
$\langle q_1|\Psi(z_1,z_2,t)\rangle$, where $|\Psi(z_1,z_2,t)\rangle$ is
given by (\ref{psifinal}).

The coincident probability of detecting photon 2 is now given by 
\begin{eqnarray}
|\langle q_1|\Psi(0,z_2,t)\rangle|^2 = {\alpha\over 2}
e^{{-2(z_2^2+z_0^2)\over\gamma^2+[{\lambda D\over\pi\gamma}]^2}}
\cosh\left[{4z_2z_0\over\gamma^2+[{\lambda D\over\pi\gamma}]^2}\right]\nonumber\\
\times\left\{ 1 + 
{\cos\left[{4z_2z_0\lambda D\pi\over\gamma^4\pi^2+\lambda^2 D^2}\right]
\over \cosh\left[{4z_2z_0\over\gamma^2+[{\lambda D\over\pi\gamma}]^2}\right]}
\right\},
\label{pattern1}
\end{eqnarray}
where $\alpha = |C_t|^2 e^{{-2z_0^2\over\epsilon^2+[{\lambda L\over\pi\epsilon}]^2}}$.
The above represents an interference pattern, even in the presence of 
the quarter-wave plates. The horizontal polarizer has erased the which-path
information for photon 1 and the interference for photon 2 has come back.
This scenario is depicted in the lower diagram of Fig.\ref{geraser}.
We call it a ghost quantum eraser because erasing the which-path information
in one photon is recovering the interference in its remote, entangled cousin,
photon 2.

One can see that the which-path information can also be erased by putting a
{\em vertical} polarizer, instead of a horizontal one. Let us see if one
recovers the interference in this case too. The vertical polarization state
can be represented by the quantum state
$|q_2\rangle=(|d_1\rangle-|d_2\rangle)/\sqrt{2}$. The state of the two photons
which reach D1 after passing through the vertical plate, and D2, respectively,
is given by
$\langle q_2|\Psi(z_1,z_2,t)\rangle$, where $|\Psi(z_1,z_2,t)\rangle$ is
given by (\ref{psifinal}).
The coincident probability of detecting photon 2 is now given by
\begin{eqnarray}
|\langle q_2|\Psi(0,z_2,t)\rangle|^2 = {\alpha\over 2}
e^{{-2(z_2^2+z_0^2)\over\gamma^2+[{\lambda D\over\pi\gamma}]^2}}
\cosh\left[{4z_2z_0\over\gamma^2+[{\lambda D\over\pi\gamma}]^2}\right]\nonumber\\
\times\left\{ 1 - 
{\cos\left[{4z_2z_0\lambda D\pi\over\gamma^4\pi^2+\lambda^2 D^2}\right]
\over \cosh\left[{4z_2z_0\over\gamma^2+[{\lambda D\over\pi\gamma}]^2}\right]}
\right\}. 
\label{pattern2}
\end{eqnarray}
The above expression represents an interference pattern which is almost exactly
the same as that in (\ref{pattern1}), except that it is shifted along the
z-axis. The shift is such that the bright fringes of (\ref{pattern2}) are
at the location of the dark fringes of (\ref{pattern1}).

\section{Discussion}

We have theoretically analysed a modified ghost-interference
setup where a which-path detector for particle 1 has been introduced.
Unravelling the particle aspect of photon 1 hides the wave aspect of photon
2. This appears to be a nonlocal extension of Bohr's complementarity principle.
We also derive a nonlocal duality relation connecting the which-path distinguishability
of particle 1 with the interference fringe visibility of particle 2.
Because of entanglement, the wave and particle aspects of the two particles
are no longer independent.

We wish to reemphasize again that this proposal should not be confused with
certain other experiments, where two-particle correlation is used to infer
the which-way information of a particle, which passes through a
double-slit\cite{kim,scarcelli,peng}.
In those experiments, the wave and particle natures in question are properties
of the {\em same} particle, only that another correlated particle is used to get
which-way information of the particle passing through a double-slit {\em and}
showing interference. In our proposal, particle 1 passes though a double-slit
and we use certain which-way detector to know which of the two slits it 
passed through. Particle 2 {\em does not pass through any double-slit},
so one cannot even talk about any which-way information for it. However,
it does show interference in coincidence with detector D1.
So the nonlocal duality relation connects which-way information for particle 1
to interference visibility for particle 2. 

We propose a realizable {\em ghost quantum eraser}
experiment. Here erasing the which-path information of one photon recovers
the interference for the other photon. The aspects discussed in this
investigation reveal highly non-classical and nonlocal features of entangled
systems.

\section*{Acknowledgments}
M.A. Siddiqui thanks the University Grants Commission, India for financial
support. Authors thank an anonymous referee for suggesting changes which
improved the clarity of the discussion.


\begin{thebibliography}{0}

\bibitem{bohr} N. Bohr, ``The quantum postulate and the recent development of
atomic theory," Nature (London) 121, 580-591 (1928). 

\bibitem{greenberger} D. M. Greenberger and A. Yasin,
``Simultaneous wave and particle knowledge in a neutron interferometer",
Phys. Lett. A 128, 391 (1988).

\bibitem{englert} B-G. Englert, ``Fringe visibility and which-way information:
an inequality", {\em Phys. Rev. Lett.} {\bf 77}, 2154 (1996).

\bibitem{ghostexpt} D.V. Strekalov, A.V. Sergienko, D.N. Klyshko, Y.H. Shih,
``Observation of two-photon ghost interference and diffraction,"
{\em Phys. Rev. Lett.} {\bf 74}, 3600 (1995).

\bibitem{ghostimaging} M. D'Angelo, Y-H. Kim, S. P. Kulik and Y. Shih,
``Identifying Entanglement Using Quantum Ghost Interference and Imaging,"
{\em Phys. Rev. Lett.} {\bf 92}, 233601 (2004).

\bibitem{rubin} S. Thanvanthri and M. H. Rubin,
{\em Phys. Rev. A} {\bf 70}, 063811 (2004).

\bibitem{zhai} Y-H. Zhai, X-H. Chen, D. Zhang, L-A. Wu,
\textit{Phys. Rev. A} {\bf 72}, 043805. (2005).

\bibitem{jie} L. Jie, C. Jing,
\textit{Chinese Phys. Lett.} {\bf 28}, 094203 (2011).

\bibitem{zeil2} J. Kofler, M. Singh, M. Ebner, M. Keller, M. Kotyrba, A. Zeiling
er,
\textit{Phys. Rev. A} {\bf 86}, 032115 (2012).

\bibitem{pravatq} P. Chingangbam, T. Qureshi,
``Ghost interference and quantum erasure,"
{\em Prog. Theor. Phys.} {\bf 127}, 383-392 (2012).

\bibitem{twocolor} D-S. Ding, Z-Y. Zhou, B-S. Shi, X-B Zou, G-C. Guo,
``Two-color ghost interference," {\em AIP Advances} {\bf 2}, 032177 (2012).

\bibitem{sheebatq} S. Shafaq, T. Qureshi, ``Theoretical analysis of two-color
ghost interference," {\em Eur. Phys. J. D} {\bf 68}, 52 (2014).

\bibitem{ghostunder}  T. Qureshi, P. Chingangbam, S. Shafaq, ``Understanding
ghost interference," arXiv:1406.0633 [quant-ph]

\bibitem{epr} A. Einstein, B. Podolsky, N. Rosen, ``Can quantum-mechanical
description of physical reality be considered complete?",
\textit{Phys. Rev.} {\bf 47} (1935) 777--780.

\bibitem{klyshko} D.N. Klyshko, ``A simple method of preparing pure states of an optical field, of implementing the
Einstein–Podolsky–Rosen experiment, and of demonstrating the complementarity principle," {\em Sov. Phys. Usp.} {\bf 31}, 74 (1988).

\bibitem{walborn} S.P. Walborn, C.H. Monken, S. P\'{a}dua, P.H. Souto Ribeiro,
``Spatial correlations in parametric down-conversion,"
\textit{Phys. Rep.} {\bf 495}, 87-139 (2010).

\bibitem{vaccaro} J.A. Vaccaro, ``Particle-wave duality: a dichotomy between
symmetry and asymmetry,"
{\em Proc. R. Soc. A.} {\bf 468}, 1065-1084 (2012).

\bibitem{tqajp} T. Qureshi, ``Understanding Popper's experiment,"
{\em Am. J. Phys.} {\bf 73}, 541-544 (2005).

\bibitem{born}  M. Born, E. Wolf, ``Principles of Optics''
(Cambridge University Press, UK, 2002), 7th edition.

\bibitem{barbosa} G.A. Barbosa, ``Quantum images in double-slit experiments with spontaneous down-conversion light," {\em Phys. Rev. A} {\bf 54}, 4473 (1996).

\bibitem{mandel} L.J. Wang, X.Y. Zou, L. Mandel, Phys. Rev. A 44, 4614 (1991); X.Y. Zou, L.J. Wang, L. Mandel, Phys. Rev. Lett. 67, 318 (1991); Z.Y. Ou, L.J. Wang, L. Mandel, Phys. Rev. A 40, 1428 (1989); C.K. Hong L. Mandel, Phys. Rev. Lett. 56, 58 (1986).

\bibitem{jaynes} E. Jaynes, in {\em Foundations of Radiation Theory and
Quantum Electronics}, ed. A. O. Barut (Plenum, New York 1980), pp. 37.

\bibitem{scully} M. O. Scully, B.-G. Englert and H. Walther, {\em Nature
(London)} {\bf 351} (1991), 111.

\bibitem{walborn1} S.P. Walborn, M.O. Terra Cunha, S. P\'adua, 
C.H. Monken, {\em Phys. Rev. A} {\bf 65}, 033818 (2002)

\bibitem{walborn2} S.P. Walborn, M.O. Terra Cunha, S.P\'adua, C.H. Monken, ``Quantum erasure",
{\em American Scientist} {\bf 91}, 336-343 (2003).

\bibitem{kim} Y-H Kim, R. Yu, S.P. Kulik, Y. Shih,
``Delayed `Choice' Quantum Eraser," {\em Phys. Rev. Lett.} {\bf 84}, 1 (2000).

\bibitem{scarcelli} G. Scarcelli, Y. Zhou, Y. Shih, ``Random delayed-choice
quantum eraser via two-photon imaging,"
{\em Eur. Phys. J. D} {\bf 44}, 167-173 (2007).

\bibitem{peng} T. Peng, H. Chen, Y. Shih,
``Delayed-choice quantum eraser with thermal light,"
{\em Phys. Rev. Lett.} {\bf 112}, 180401 (2014).

\end{thebibliography}
\end{document}